# Using Side Channel Information and Artificial Intelligence for Malware Detection

Paul Maxwell
*Army Cyber Institute*
United States Military Academy
West Point, NY
paul.maxwell@westpoint.edu

David Niblick
*Dept. of Electrical Engineering and Computer Science*
United States Military Academy
West Point, NY
david.niblick@westpoint.edu

Daniel C. Ruiz
*Dept. of Electrical Engineering and Computer Science*
United States Military Academy
West Point, NY
daniel.ruiz@westpoint.edu

*Abstract*—Cybersecurity continues to be a difficult issue for society especially as the number of networked systems grows. Techniques to protect these systems range from rules-based to artificial intelligence-based intrusion detection systems and anti-virus tools. These systems rely upon the information contained in the network packets and download executables to function. Side channel information leaked from hardware has been shown to reveal secret information in systems such as encryption keys. This work demonstrates that side channel information can be used to detect malware running on a computing platform without access to the code involved.

## I. INTRODUCTION

Computer systems of all types have been subject to attacks almost since their inception. Attackers seek to steal information, obtain monetary benefits, prevent access to systems, or simply cause aggravation for the users. As a result, research and commercial products have emerged to counter these attacks both on the target system and at network entry points. These Intrusion Detection Systems (IDS) and Anti-Virus (AV) suites attempt to protect the user from harm. These systems often use methods such as signature recognition, rule matching (behavior-based, heuristic-based), and artificial intelligence (AI) models to do their work[1]. Accordingly, these tools need access to the network packets or code running on a machine to function. Often, these tools require static analysis of the malware to perform their function. As a result, keeping up with the latest malware and attack methods for signature and rules-based systems is a challenge due to constantly evolving techniques and obfuscation tools that change the signature of malware binaries. AI-based methods rely upon subject matter experts to pre-process the information used for detection (features) and potentially to constantly update their models as new attacks are created. These factors make protection product development challenging.

The growing research field of Side Channel Analysis (SCA) has demonstrated that hardware devices can leak information that can be used to discover secrets about a system. Successful side channel attacks are capable of breaking encryption keys in seconds with only a minimal knowledge required of the underlying encryption algorithm used. These side channels can take the form of voltages, magnetic fields, optical waves, audio signals, and more. The underlying principle that allows side channel analysis to be successful is that hardware system emissions can depend upon the operations being performed by the executing software code. This dependency allows attackers to reverse engineer systems to determine what instructions are executing and therefore learn pieces of secret information such as encryption key bits.

In this work, we develop a proof-of-concept system to demonstrate that side channel information from computing systems can be used to detect malware execution. Information readily available from the hardware such as core temperatures, fan speeds, and memory usage provide data that when analyzed by artificial intelligence models reveal the presence of malware on a system. The contributions of this work are proof-of-concept models for detecting malware without access to the code running on the system and an initial dataset to support research in this area. A tool of this type can potentially simplify the protection of systems as detection of obfuscated or encrypted code is overcome. Malware will still perform the same functions on the hardware despite minor changes to its structure. The potential to use this tool in cloud or zero-trust environments to protect clients without access to their software and data is enormous.

The remainder of this paper is structured as follows. A synopsis of related work is provided in Section II. Section III discusses the design of the AI models used and how the dataset was created. Section IV presents the results of the experiments and Section V provides a conclusion and recommendations for further research work.

## II. RELATED WORK

There are numerous research efforts that use artificial intelligence models to detect network intrusions [2]–[5]. These products utilize a variety of models to accomplish their goal, but all rely upon access to the contents of network packets or system information such as the output logs from other tools (e.g., audit logs, Intrusion detection system output). The packets and other sources are parsed and processed to provide inputs to the AI systems. Often these systems rely on subject matter experts to pre-process the raw data features and to choose the applicable features and how to best represent those features thus limiting their usefulness to the expertise of the human designer. Research in this area also derives many of its techniques from the image processing realm which is not always effective. As Maxwell et al. point out [6], there is more work to be done to determine the best pre-processing techniques for network data.

Further limiting the usefulness of AI-based approaches is the increasing application of adversarial AI. These techniques can cause the malfunction of AI systems through data poisoning, creation of altered inputs, and more [4]–[6]. Systems that rely upon data that can be altered by an adversary are subject to reduced effectiveness and reliability.

Research into applications that harvest information from leaked side channel data has produced many results. Side channel information has been used to attack cache contents [7], memory contents [8], and to break encryption keys used in systems [9]. These systems use the properties of the target hardware to elicit information. This is possible due to the correlation between actions in executing code, such as performing floating point multiplication that is only in a conditional section of code, and physical manifestations of that action such as increased current draw, changes in a magnetic field, or higher fan speeds to cool the chip. This information leaked via these physical properties can then be processed using mathematical techniques to discover the desired secret.

Side channel information has been used in some areas for anomaly detection such as in industrial control systems. The work in [10] uses the execution times of uninfected golden systems and then compares these times to detect altered deployed systems. Their technique uses knowledge of the base system and a p-test to detect the alterations. The research performed in [11] attempts to detect incorrectly operating industrial control systems using electromagnetic (EM) emissions. Using a baseline measurement and real-time EM emissions a comparison is performed using statistical techniques to detect anomalies. Methods such as these demonstrate the potential of side channels to reveal intrusion, but they still rely on an understanding of the underlying software running on the target machines and a definition of what is 'normal'. Additionally, these systems require sensing hardware that is in close proximity to the protected device whereas our proposed solution can execute using only a software package running on the target machine.

Work using voltage levels and battery usage to detect anomalies on mobile computing systems has been done in works such as [12], [13]. Here baseline battery usage was determined a-priori and then compared to real-time usage to detect unauthorized software. This lightweight method was somewhat effective on the simpler mobile devices of the time but still required the baseline assessment of a non-infected system and some expert knowledge.

The researchers in [3] use a Multi-Layer Perceptron (MLP) network to analyze electromagnetic emissions for intrusion detection in cyber physical systems. In this work, the AI model is trained on a system's EM emissions while in a known clean state and then run while the system is deployed to detect anomalies. As with the industrial control systems, this system requires external monitoring devices and the capture of a baseline for each protected device. As new software is added to a system, this baseline would need to be updated and the model retrained to make it useful. Our system would not require updates to the detector when the user makes changes to the overall software environment.

Wang et al. [14] use LSTM, B-LSTM, MLP, and autoencoders to perform anomaly detection in Programmable Logic Controllers and Arduinos using measured power supply voltages. Using their method, they were able to detect anomalies in these systems, though with a very small test set. This technique is limited to one side channel and normally requires additional measurement tools to capture the voltages in question. Additionally, modern systems protect voltage controllers in such a way that makes measurement of voltages difficult and newer chips are developing technologies that maintain consistent voltages regardless of the operation using techniques such as Dual-Rail Pre-charge circuits [15]. Our technique utilizes many side channel sources and does not require altering the protected hardware system.

III. DATASET AND MODEL DESIGN

A. Dataset Creation

To our knowledge, no dataset currently exists that contains side channel information of clean and malware infected systems. One method of training AI systems is to use supervised learning methods. This requires a labelled dataset to achieve its goals. As a result, we created a dataset of side channel information to use in this study, which is one contribution of this work. The dataset can be found in IEEE's dataport dataset hosting site [16].

The method used to gather the training and testing data for our system relied upon a free-ware software tool named HWiNFO [17]. This real-time monitoring tool can be installed on most operating systems and can report various data points about the hardware such as: memory load, core temperatures, core frequency, core usage percentage, and write/read rates. The data reported is selectable by the user and for this work, we used 132 features from the software's output. This software runs in the background and reports its measurements to the display (optionally), and if desired, to a .csv file at a selected sampling rate (2 samples per second for this work). The output of this tool is then used as the input to our AI models for the purpose of classifying processor activity as either benign or malicious.

Many sources exist from which malware code can be obtained. Two sources of malware used in this work are *theZoo* repository on GitHub [18] and *VirusShare* [19]. These repositories were chosen due to their ease of use and their free access to the source executables. Additionally, they provided malware that was likely to run on the targeted operating systems. Malware was used from these sites that through forensic evidence we were able to verify its activity. This involved researching each piece of potential malware for its signatures and forensic artefacts and then searching for those artefacts on the target machine after malware execution. This proof-of-execution requirement limited the number of malware samples that we could generate.

The current dataset has 57 samples. A breakdown of the dataset contents is provided in Table 1. This is an admittedly small dataset, but due to the challenge of creating the dataset for this proof-of-concept work, we believe it is a good starting point. In this set there are 29 malware samples and 28 benign samples. Each sample is a time series output from the HWInfo capture lasting approximately eight minutes. Samples were gathered from six different hardware PCs (standard laptops and desktops) using VMWare/VirtualBox virtual machines running one of two different operating systems (Win7SP1 and Win XP Pro). Malware sample files were created such that the malware executable began execution at randomly selected times from the set {90, 120, 150} seconds. This was done to avoid the models from simply detecting malware execution start times. Before the malware execution start time, benign software was run to simulate normal behavior. All time-series samples were hand-labeled by row to indicate the status of the machine at that point in time (benign, malicious). In our current dataset, each file is named as follows: benign/malware name; operating system; hardware system id number; category of malware; and start time of malware execution. The intent of this naming convention is to allow future research to be performed on topics such as transferability of the models. The malware comes from one of the following categories: ransomware, worms, trojan backdoors, and spyware. Benign samples were generated by users executing routine software such as web browsers, word processing software, games, system maintenance tools, or the benchmarking software tools such as PCMark 5 and PCMark7 [20].

Table 1. Dataset Sample Composition.

| File Classification | Executable Type | Samples in Dataset |
|---|---|---|
| Benign | OS-only | 6 |
| Benign | Pcmark Test Suites | 12 |
| Benign | Simple games | 2 |
| Benign | Complex code | 1 |
| Benign | Office-type software | 4 |
| Benign | System tools | 3 |
| Malicious | Ransomware | 11 |
| Malicious | Worms | 9 |
| Malicious | Spyware | 1 |
| Malicious | Trojan back doors | 6 |
| Malicious | Virus | 1 |
| Malicious | Rootkit | 1 |

*B. Model Design*

A Multi-Layer Perceptron model from the Scikit-Learn API [21] was used as a baseline of comparison for this work. This model was trained using a *tanh* activation, an *adam* solver, an adaptive learning rate, and a maximum of 1000 iterations.

Because the samples in the dataset are time series and the information contained in the rows are correlated, we chose to build a model that would attempt to find meaning over different sampling periods. In addition, some malware has activities that are a function of time such as time-based beaconing that contacts a remote server to receive instructions. To detect this type of activity, simple analysis of a single row in the sample file is insufficient. One of our models gained inspiration from Granat [22] who combined a time series signal with smoothed (i.e., window averaged) and down-sampled versions of the signal as input to a classifier. The applications for their source model were time series applications such as stock market prediction or detection of anomalous heart rhythms.

For our work, the original signal is provided as one of several inputs to the keras Convolutional 1-dimensional network as shown in Figure 1. This network uses 64 filters, a kernel size of 32, and a *tanh* activation. It uses a GlobalMaxPooling1D layer which then feeds a Dense layer that uses a *tanh* activation, a L1_L2 regularizer, and an L2 activity regularizer. A dropout layer of 0.3 is used before the final Dense layer that uses a *sigmoid* activation. In addition to the original signal, we provide two time series inputs with rolling window averages (smoothed signals) of different lengths (2.5 seconds and 12.5 seconds). These smoothed signals attempt to filter out unwanted noise from the signal. Finally, we provide two time series inputs down sampled to look for periodic signals with a longer period (5 seconds and 12.5 seconds) than the original signal. Experiments were performed on each of the model variables to include the averaging and sampling windows to find the best setting.

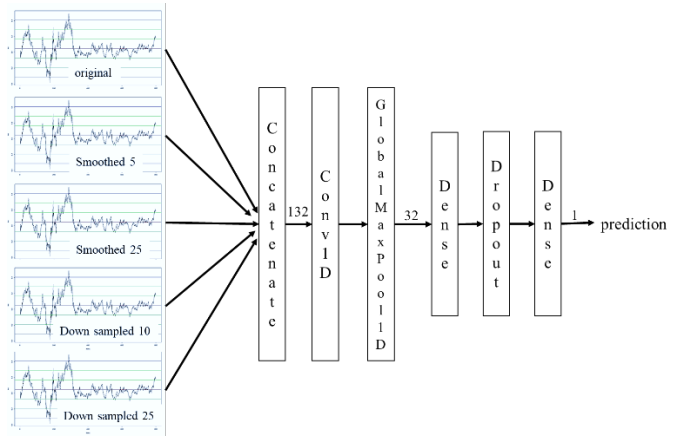

Fig 1. One Dimensional Convolutional Neural Network Classifier Architecture.

Unlike many AI classifiers that perform predictions based upon a single row of data, this system uses an entire input sample file to perform its classification. The use case for this system is for the model to run in the background of a system using the side channel sampling software as input. If the system is uncompromised, the model would continuously report a benign state. Once malicious activity is detected, the model would report a malicious state until the system is cleaned of the malware. Because most of the under-lying models do make predictions for each row/time sample (except for the Recurrent Neural Network models), an additional classifier routine was added to the models. This routine establishes a threshold of consecutive samples classified as malicious that result in the file being classified as malicious. Through experimentation, we set this threshold at 50 samples. The relationship between this threshold and the model accuracies is shown in Figure 2. Adding this threshold has the benefit of preventing transient errors from misclassifying the file (i.e., a false positive).

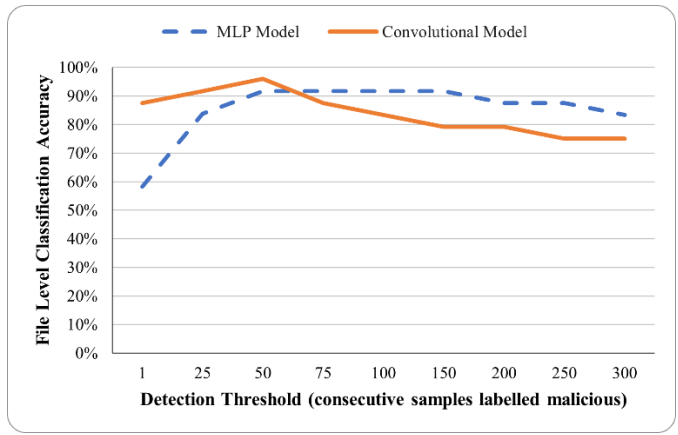

Figure 2. Detection Threshold versus File Level Accuracy

Given the nature of the dataset, noise was a particular concern. Autoencoders have been shown to be particularly successful at denoising data [23], [24]. Autoencoders work by reducing the dimensionality of the data while minimizing the loss of information. This is accomplished by training a model such that both the input and output are the same signal. However, the hidden layers are purposely constricted. Forcing the model to recreate the same signal, but with fewer nodes in the hidden layers causing the data to be represented by fewer dimensions in those hidden layers. In theory, the "important" information remains at the output, but the "noise" is removed. Once trained, we remove the output layer (i.e., decoder layer), and use the encoding layer with its reduced dimensions as input to the models. This is the denoised "encoded" signal.

We trained seven MLP-based autoencoders, reducing the dimensionality of the hidden encoding layer to 5, 10, 15, 20, 30, 40, and 50 dimensions, respectively. Once trained, we encoded our dataset to those dimensions. The encoded datasets are essentially denoised to varying degrees. We repeated the experiments on the encoded datasets to compare against each other and the original results. The effects of the encoder's dimensionality on file classification accuracy is shown in Figure 3.

Furthermore, given that our dataset is inherently comprised of time-series data, we naturally sought to implement an alternative set of models specifically tailored to extract meaning from sequences. As a result, we trained five separate recurrent neural networks (RNNs) to experimentally determine an acceptable approach. RNNs have been shown to perform well on time series data in a variety of environments [25]–[27]. RNNs are deep neural networks that utilize specialized neurons to maintain an internal memory state. This mechanism enables such networks to learn from the position and order of individual values within a sequence, rather than considering each value in isolation. We first created a simple RNN with standard recurrent neurons consisting of four hidden layers, each comprised of 16, 32, 32, and 16 neurons. These models were trained using *tanh* activation and an adaptive learning rate, albeit with an *RMSprop* optimizer and a maximum of 100 epochs. Each RNN we trained utilized this architecture, with the primary difference derived from the types of recurrent neurons used, to include gated recurrent units (GRUs) and long-short term memory cells (LSTMs), and the incorporation of bi-directionality. In total, this network configuration contained approximately 6000 trainable parameters, which we estimated would result in efficient training while still enabling the network to generalize. Ultimately, our various RNN models performed almost identically to one another when used to classify sequences. However, while implementing these RNNs, the length of each input sequence emerged as a crucial hyperparameter, one which yielded notable results when used to classify sub-samples of our dataset's original samples. The RNN architecture is shown in Figure 4.

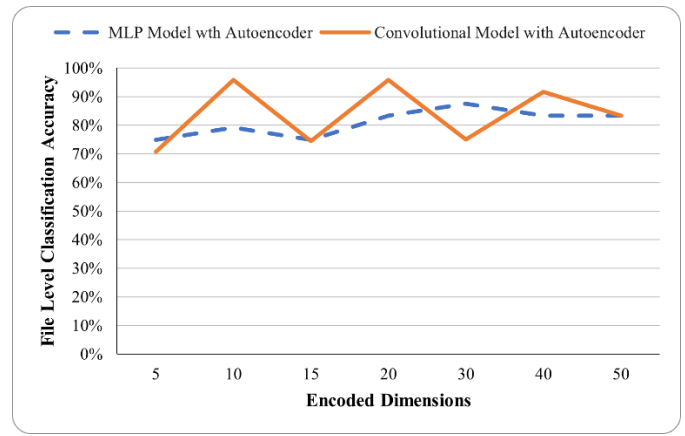

Figure 3. File Level Classification Accuracy versus Encoding Dimensions used in Autoencoder.

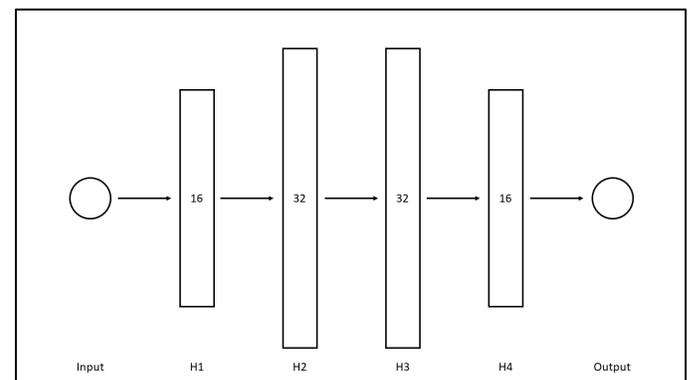

Figure 4. Recurrent Neural Network Architecture.

## IV. MODEL RESULTS AND DISCUSSION

The Convolutional, MLP, and autoencoder models were trained and tested on standard laptops running Ubuntu Linux and Windows 10 using Python3. The dataset was separated into training and testing subsets with 32 file samples in the training set (16 benign, 16 malicious) and 24 file samples in the test set (11 benign, 13 malicious). For the RNN model, sequences of varying length (5-960) were created from the dataset by breaking the input files into the appropriately sized chunks. Each sequence was then used as an input to the RNN models for training and testing. The members of each subset were selected to have representatives from each malware category and similar numbers of samples with equivalent malware execution starting times. During training, the files were shuffled to randomize their order during training. Each file was provided to the model for training using a generator function. For testing, the input samples were not shuffled.

The accuracy of each of the models is shown in Table 2. The row accuracy is the classifier's accuracy based on each input row (i.e., time sample) from the sample file. Because RNNs use sequences of data, a single sample accuracy is not applicable and is thus not reported. The file sample accuracy is the file-level accuracy for the sample. In other words, an input sample file that is benign should have a file sample classification of benign while a malicious sample would have a single file sample malicious label. Also shown are the false positive rates and false negative rates for the tested models based on the individual samples for the Convolutional and MLP models and the file samples for the RNN models. Both metrics are important for this application domain. False positives create alarms when no malware exists thereby creating work for IT staff in resolving the alarms and loss of user productivity. False negatives are more dangerous in this environment as they represent undetected malware. Obviously, the desire is, in an ideal detector, to have zero false negatives and as few false positives as possible. In this work, we can see that the convolutional network performed better than the other models in all metrics. Modern detection systems currently have better performance than our models but these results show the promise of this technique with the benefit of requiring no knowledge of the malware code nor access to the system other than to run the sampling tool. Further research in this area should result in improved models that can approach or exceed current tool capabilities.

Using an autoencoder to denoise data prior to training a model has mixed results. At best, it matches performance of the original implementation, and at worst, it significantly degrades performance. The 1D Convolutional model obtains similar performance with a reduced input dimensionality which is important in developing smaller, faster models and models that are more robust against adversarial perturbation. Future work can be done to identify the most important features in the dataset. Additionally, there is no single encoded dimensionality that optimizes performance across both the MLP and 1D Convolutional models. This implies that there is not an optimal amount of denoising inherent to the dataset, but that if denoising is used, it must be tweaked specific to the model and process. This adds another layer of complexity to the process without much benefit.

Table 2. Model Accuracy, False Positive Rates, and False Negative Rates

| Model | Data Sample (row) Accuracy | File Sample Accuracy | False Positive Rate | False Negative Rate |
|---|---|---|---|---|
| Multi-Layer Perceptron | 83.82% | 91.67% | 5.15% | 14.14% |
| Convolutional 1D | 85.47% | 95.83% | 1.66% | 1.66% |
| Multi-Layer Perceptron with Autoencoder | 69.52% | 83.33% | 8.90% | 20.17% |
| Convolutional 1D with Autoencoder | 85.10% | 95.83% | 0.55% | 18.76% |
| Recurrent Neural Network | - | 90.91% | 9.00% | 0.00% |
| LSTM RNN | - | 90.91% | 9.00% | 0.00% |
| LSTM RNN (Bidirectional) | - | 90.91% | 9.00% | 0.00% |
| GRU RNN | - | 81.82% | 9.00% | 9.00% |
| GRU RNN (Bidirectional) | - | 90.91% | 9.00% | 0.00% |

Regarding the demonstrated performance of our RNN models, an important limitation to mention is the fact that these models were trained and evaluated on fewer file samples than their counterparts. Due to the need to define a constant sequence length, the RNN models in the Table 3 ignored any file samples consisting of less than 960 rows (8 minutes, equivalently). Thus, the RNNs were trained using only 16 of 32 available samples and evaluated using 11 of 24 available test samples. Despite these limitations, our RNN models performed reasonably well, which leaves open the potential for improved accuracy, given a larger dataset with a more even distribution of sample lengths.

*Table 3. RNN Model Accuracy on Test Data for Different Sequence Lengths*

| Sequence Length (500ms) | Training Sequences | RNN | LSTM RNN | Bi-directional LSTM RNN | GRU RNN | Bi-directional GRU RNN |
|---|---|---|---|---|---|---|
| 5 | 6469 | 99.39% | 99.39% | 99.39% | 99.39% | 99.39% |
| 20 | 1606 | 97.55% | 97.34% | 97.55% | 97.55% | 97.55% |
| 40 | 794 | 95.01% | 95.01% | 95.01% | 95.01% | 95.01% |
| 80 | 389 | 89.74% | 89.74% | 89.74% | 89.74% | 89.74% |
| 160 | 189 | 78.95% | 79.82% | 78.95% | 78.95% | 78.95% |
| 320 | 84 | 68.00% | 66.00% | 66.00% | 66.00% | 66.00% |
| 640 | 38 | 52.38% | 57.14% | 71.43% | 66.67% | 66.67% |
| 960 | 16 | 90.91% | 90.91% | 90.91% | 90.91% | 90.91% |

More interestingly, while exploring the application of RNNs to this problem, we used the sequence length hyperparameter to subsample our dataset to varying degrees. This allowed us to artificially increase the number of samples in our dataset. The performance of each model across a range of sequence lengths is given in Table 3 below. Notably, all recurrent models performed equally well on sub-samples of length 80 or lower, and even exceeded the sample classification accuracy of our convolution models for sub-samples with a length of 40 or lower. Additionally, our bidirectional LSTM network outperformed its peers for subsamples of length 640. These results suggest that RNNs could serve as the primary backbone in live malware detection systems that process side-channel data in shorter intervals. It is our hypothesis that the smaller sequences have a higher probability of possessing uniform labels and thus the model is better able to classify these sequences. As the sequence size grows, the probability of having mixed labels increases and the model struggles to perform as well. This is shown to improve though as the sequence continues to grow and the model ingests more data for its analysis. It may be possible that this fine-grained approach to side channel monitoring and classification could lead to earlier alerts of malicious activity on a given system, and certainly warrants further research.

The time to detect the malware by the convolutional and MLP models is shown in Figure 5. The time to detect was calculated by determining the difference between the start time of malware execution and the time at which the model flagged the presence of malware. On average, the MLP model detected malware more quickly than the convolutional model though it failed to detect malware more frequently. The average time to detection for the MLP model is 54.33 seconds while the Convolutional model took 55.19 seconds on average. Given that the malware detection threshold in our tool is 25 seconds, these values are not too high for practical use.

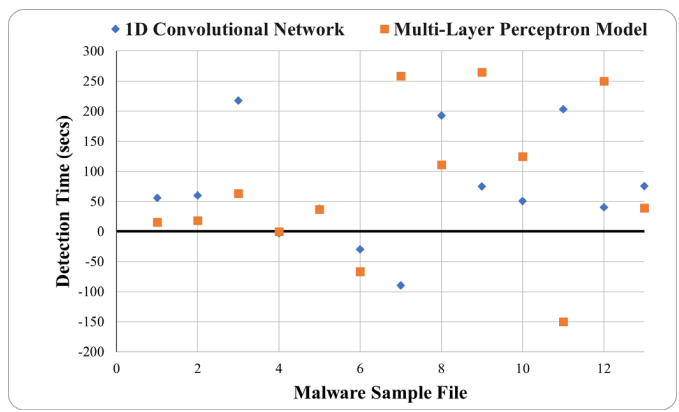

**Figure 5.** Malware detection times by malware file sample for each model.

V. CONCLUSIONS AND RECOMMENDATIONS.

This work has demonstrated a proof-of-concept malware detection system that relies upon side channel information from hardware systems. Without accessing source binaries or network packets, we detected the presence of active malware with greater than 91% accuracy using a simple MLP model, 95.83% accuracy using a Convolutional model, and 90.91% using a Recurrent Neural Network. Furthermore, our various RNN models perform well on file sample classification despite training on less data, and they outperform our Convolutional model for shorter sequence lengths. A system using these techniques can be invaluable to cloud providers and others who do not have permissions on the target machine. In addition to this contribution, we have created a malware dataset that can be expanded and used by other researchers for work in this area. Datasets such as these are the foundation for many research efforts in artificial intelligence applications.

Future work in this area includes growing the dataset to include more malware samples by variety and complexity along with samples running on more diverse hardware architectures and operating systems. Additional research should be done to determine the extent to which transferability can be achieved. This is critical to making this system broadly useable without tuning it to a specific machine or operating system. Research can also be done to determine if the models can classify malware by its type instead of just detecting its presence. Finally, work can be done to identify which features from the side channel data tool are important so that the dimensionality of the data can be reduced.